\documentclass[12pt]{article}
\usepackage{amssymb}
\usepackage{amsfonts}
\usepackage{graphicx}
\usepackage{epsfig}

\newcommand{\ini}{\begin{equation}}
\newcommand{\barray}{\begin{eqnarray}}
\newcommand{\fin}{\end{equation}}
\newcommand{\earray}{\end{eqnarray}}
\newcommand{\equ}[1]{~Eq.(\ref{#1})}

\newcommand{\bmath}{\begin{displaymath}}
\newcommand{\emath}{\end{displaymath}}
\newcommand{\bite}{\begin{itemize}}
\newcommand{\eite}{\end{itemize}}

\newcommand{\bx}{{\bf x}}

\newcommand{\tr}{{\rm Tr}}
\renewcommand{\o}{\omega}
\newcommand{\m}{\mu}
\newcommand{\n}{\nu}
\newcommand{\eps}{\varepsilon}

\begin{document}

\psfull

\vspace{0.5cm}
\begin{center}
{\Large \bf A Center-Symmetric 1/N Expansion}\\
\vspace{0.7cm}
\vspace{0.5cm} {\large
Martin Schaden\footnote{email: mschaden@andromeda.rutgers.edu}}\\

\vspace{1.5cm}
{\it Department of Physics, Rutgers University in Newark,\\
365 Smith Hall, 101 Warren Street, Newark, NJ 07102, USA.} \\
\end{center}

\bigskip

\begin{center}
\bf Abstract
\end{center}
The free energy  of $U(N)$ gauge theory is expanded about a
center-symmetric topological background configuration with
vanishing action and vanishing Polyakov loops. We construct this
background for $SU(N)$ lattice gauge theory and show that it
uniquely describes center-symmetric minimal action orbits in the
limit of infinite lattice volume. The leading contribution to the
free energy in the $1/N$ expansion about this background is of
${\cal O}(N^0)$ rather than ${\cal O}(N^2)$ as one finds when the
center symmetry is spontaneously broken. The contribution of
planar 't Hooft diagrams to the free energy is ${\cal O}(1/N^2)$
and sub-leading in this case. The change in behavior of the
diagrammatic expansion is traced to Linde's observation that the
usual perturbation series of non-Abelian gauge theories suffers
from severe infrared divergences\cite{Linde80}. This infrared
problem does not arise in a center-symmetric expansion. The 't
Hooft coupling $\lambda=g^2 N$ is found to decrease $\propto
1/\ln(N)$ for large $N$. There is evidence of a vector-ghost in
the planar truncation of the model.
\bigskip


\newpage

\section{Introduction}

Confinement can be defined as the absence of asymptotic states in
non-trivial multiplets of the global gauge group. Since the number
of singlet states does not increase proportional to $N$, the free
energy of a $U(N)$ gauge theory in a confining phase should be of
order $N^0$\cite{Pisarski84}. Perturbatively, the adjoint
multiplet of gauge bosons and the fundamental fermion multiplets
contribute to the free energy density of $U(N)$ gauge theory in
${\cal O}(N^2)$ and ${\cal O}(N)$. A direct application of 't
Hooft's $1/N$ expansion\cite{THooft74a,Witten79} apparently also
gives a free energy density of order $N^2$ even at low
temperatures. Using the Dyson-Schwinger equations of the lattice
and Migdal's factorization condition for planar diagrams, Gocksch
and Neri\cite{GoNe83} on the other hand found that the free energy
density in the confining phase does not depend on the temperature
at $N=\infty$.

A leading contribution to the free energy of $U(N)$ gauge theory
of order $N^2$ is not compatible with the result of Gocksch and
Neri\cite{GoNe83}. It is more reasonable to assume that the
coefficient of the $N^2$-term in the $1/N$-expansion of the free
energy vanishes in the confining phase and that the model defined
by planar diagrams is a topological theory without dynamical
degrees of freedom. It turns out that this is the case in the
$N\rightarrow\infty$ limit only. The factorization for large $N$
in the confining phase implies that $U(N=\infty)$ is described by
a matrix model that depends on space-time
parametrically\cite{EK,QEK,TEK,Das87} only.

The confining phase of pure Yang-Mills models is characterized by
a global center symmetry. This symmetry also is essential in the
formulation of reduced models\cite{QEK,TEK,Das87} at $N=\infty$.
The objective here is to construct an $1/N$ expansion that
preserves this center symmetry in every order. Since the pure
Yang-Mills action is invariant, one can achieve this by expanding
about a center-symmetric topological field configuration.

We show the absence of contributions to the free energy
proportional to $N^2$ and $N$ when $SU(N)$ gauge theory is
expanded about such a center-symmetric orbit.  In this expansion,
planar 't Hooft diagrams contribute to the free energy in order
$1/N^2$. The present analysis systematizes and extends the result
of Gocksch and Neri\cite{GoNe83} in several ways. The
center-symmetric $1/N$ expansion is possible for all $N$. It not
only gives the order of planar contributions to the free energy
but also of higher genus 't Hooft diagrams. Although fields in the
fundamental representation explicitly break the center symmetry,
there are no contributions to the free energy of order $N$ in this
expansion -- planar diagrams with a single fundamental color loop
contribute in ${\cal O}(1/N^3)$. The leading temperature dependent
contributions to the free energy are of order $N^0$. These
non-planar contributions survive the large $N$ limit -- as might
be expected if the masses of asymptotic singlet states have a
finite limit\cite{Witten79}.

In the topological sector with vanishing instanton number that
interests us here, the classical action of a gauge theory vanishes
at field configurations with minimal action.  However, this
classical field is not necessarily a pure gauge configuration.
Since the local curvature of the configuration vanishes, the
possibly nontrivial gauge invariant quantities are
non-contractible Wilson loops. [These non-contractible loops in
general are sensitive to global symmetries of the action and thus
can distinguish different phases of the model.] At a finite
temperature $T$ and infinite volume ${\cal V}$, configurations
with vanishing curvature are characterized by their Polyakov
loops, non-contractible Wilson loops in the Euclidean temporal
direction.

Specifically, consider the Polyakov loop of an $SU(N)$ gauge
theory at finite temperature $T$ with periodic boundary conditions
for the connection,
\begin{equation}\label{Ploop}
{\cal L}({\bf x})=\tr U(x)\ \ {\rm with} \ U(x)={\cal
P}\exp\left[i \int_{x_4}^{x_4+1/T} V_4({\bf x},\tau) d\tau\right]\
.
\end{equation}
In \equ{Ploop} ${\cal P}$ denotes ordering of the exponential
along the path and $V_\m$ is the gauge connection in the
fundamental representation. On the lattice, $U(x)$ is the ordered
product of the links in the periodic temporal direction, beginning
with the link at $x$. One can choose a gauge in which $V_4({\bf
x},\tau)$ does not depend on the Euclidean time $\tau$ and is
diagonal. On the lattice this may be achieved in three steps: one
first uses the gauge freedom to set all temporal links apart from
those on the $x_4=0$ time slice to unity. [The nontrivial temporal
links of this representative configuration then are the $U({\bf
x},x_4=0)$ of the Polyakov loop.] One next uses time-independent
gauge transformations to diagonalize the remaining nontrivial
temporal links. Since the permutation group is a subgroup of
$SU(N)$, the phases in addition can be ordered so that the
temporal links of the $x_4=0$ time slice are of the form,
\begin{eqnarray}\label{diagonalU}
U(\bx,x_4=0)&=&{\rm
diag}(e^{i\theta_1(\bx)},\dots,e^{i\theta_N(\bx)}),\
\ {\rm with}\ \sum_{j=1}^N\theta_j(\bx)=0\ ,\nonumber\\
&&\hspace{-2em}{\rm and}\
-\pi\leq\theta_1(\bx)\dots\leq\theta_j(\bx)\leq\theta_{j+1}(\bx)\dots\leq\theta_N(\bx)<\pi\
,
\end{eqnarray}
and all other temporal links are unity. The Abelian invariant
subgroup of the configuration is enhanced to a non-Abelian one
when some of the phases in\equ{diagonalU} are degenerate. The
corresponding continuum configuration in this case may have a
non-trivial monopole number\cite{THooft79}. [Since all lattice
configurations are contractible, the usual topological
classification of smooth continuum configurations cannot be used,
but degenerate configurations that are invariant under a
non-Abelian subgroup of $SU(N)$ can also be found on the lattice.]

One finally may use time-dependent Abelian gauge transformations
to evenly distribute the $U(\bx,0)$ of\equ{diagonalU} in temporal
direction. In the continuum limit, the resulting configuration
corresponds to a temporal component of the connection $V_4(\bx)$
that does not depend on the Euclidean time $x_4$ and is Abelian.

The perturbation series can be constructed about any configuration
of minimal classical action. Although such a configuration
generally will not correspond to a minimum of the effective
action, the perturbation series nevertheless yields some
information about the configuration space in its vicinity. We will
see that the confining phase extends to arbitrary small values of
the 't Hooft coupling for sufficiently large $N$. That the
effective coupling may become weak in the confining phase at
sufficiently large $N$ was previously observed\cite{Polchinski93}
by exploiting the analogy with string theory. We show that the
perturbative analysis leads to the same conclusion.

In the topologically trivial sector, the local curvature of a
minimal action orbit vanishes. The previous construction implies
that the temporal links of periodic lattice configurations with
{\it minimal} Wilson action can be chosen Abelian {\it and}
constant across the whole lattice. Since every plaquette-action of
a minimal action configuration vanishes and all temporal links
apart from those on a particular time slice can be set to unity by
a gauge transformation, the spatial links of a minimal action
configuration do not depend on time in such a gauge. Periodicity
of the configuration in time then requires that the eigenphases of
two spatially adjacent temporal Abelian links are the same: since
{\it all} plaquette-actions vanish we must also have that $g
a=a'g$, or $g a g^\dagger=a'$ for two equal spatial links $g\in
SU(N)$ and two adjacent temporal links $a$ and $a'$ on the $x_4=0$
time-slice. The previous procedure shows that $a$ and $a'$ can be
chosen to lie in the Abelian subgroup of $SU(N)$. $a$ and $a'$
thus are the same up to a permutation of their eigenphases. Taking
into account that the eigenphases have been ordered, one concludes
that $a=a'$ in this particular gauge.

All temporal links on the $x_4=0$ time slice of this
representative of an orbit with minimal Wilson action thus are
Abelian and the {\it same} -- all other temporal links are unity.
We in particular have that minimal action configurations of a
time-periodic $SU(N)$-lattice are characterized by a Polyakov loop
that does not depend on the chosen spatial point. A spatially
constant Abelian gauge transformation can be used to evenly
distribute the temporal links of the $x_4=0$ time-slice in
temporal direction. One thus obtains a representative of any orbit
with {\it minimal} Wilson action that is described by a temporally
and spatially constant Abelian connection $V_4$. When none of the
eigenphases of the temporal links are degenerate, spatial links in
this gauge also have to be in the Abelian subgroup and do not
depend on Euclidean time.

\section{Topological configurations and center symmetry}
The minimal action configurations of $SU(N)$ are further
characterized by their transformation under a global $Z_N$
symmetry of the Wilson action. This so-called center symmetry is
generated by multiplying every temporal link on a particular time
slice by an element of the center of $SU(N)$ -- possibly followed
by a (periodic) gauge transformation of the configuration. This
transformation multiplies the Polyakov loops of any configuration
by a root of unity, but does not change the Wilson action. The
center symmetry therefore maps minimal (Wilson) action
configurations onto themselves. It allows to distinguish between
minimal action orbits that are invariant under this discrete
global symmetry and those that are not.

\subsection{The center-symmetric topological configuration}
Since any Polyakov loop  is  multiplied by a root of unity, an
orbit is center-symmetric only if its Polyakov loops vanish. The
$N$ eigenphases of $U(x)\in SU(N)$ in\equ{Ploop} therefore sum to
zero and their product is ${\rm det} U(x)=1$. The discussion in
the introduction shows that one may choose $U({\bf x},0)$ constant
and in the Abelian subgroup. The constant $\theta_j^{(0)}$ on the
$x_4=0$ time slice of such a center-symmetric minimal action
configuration thus are,
\begin{equation}\label{symvac}
\theta_j^{(0)}({\bf x})=\pi (2j-N-1)/N\,,\ {\rm for}\
j=1,2,\dots,N\ .
\end{equation}

A center transformation simply permutes the phases in\equ{symvac}
and the previous ordering can be restored by a time-independent
$SU(N)$ gauge transformation (of which the permutations are a
subgroup). The fact that the eigenphases in\equ{symvac} are
equidistant was recently exploited to define an order parameter
for the center-symmetric phase\cite{Neuberger03}.

None of the eigenphases of a center-symmetric configuration with
minimal action are degenerate. The spatial links therefore do not
depend on time and are Abelian as well. On a lattice that is
periodic in every direction they can in fact be chosen Abelian and
constant. To see this, one may proceed as follows.  Using
time-independent Abelian gauge transformations only, all (already
time independent) spatial links in $x_3$-direction apart from
those on the $x_3=0$ slice may be set to unity. This
time-independent Abelian gauge transformation does not change the
temporal links. Since this is an Abelian minimal action
configuration on a lattice that is periodic in $x_3$, the links in
the $x_3$-direction on the $x_3=0$ slice in fact must all be
equal. The remaining Abelian links in the $x_2$- and
$x_1$-directions at this stage do not depend on $x_3$ (nor on
$x_4$). Using an Abelian gauge transformation that depends on
$x_3$ only, the links in $x_3$-direction on the $x_3=0$ slice can
be distributed evenly in the $x_3$-direction. The result is a
gauge equivalent configuration with constant Abelian links in
$x_4$- and $x_3$- directions and Abelian links in $x_2$- and
$x_1$- directions that do not depend on $x_3$ nor on $x_4$. The
procedure is repeated with $x_4$ and $x_3$-independent Abelian
gauge transformation to also make the links in $x_2$-direction
constant (links in $x_1$-direction at this point do not depend on
$x_2, x_3$ nor $x_4$). Abelian gauge transformations that depend
only on $x_1$ can finally be used to obtain a configuration with
Abelian links in each direction that do not depend on space or
time.

In general there are inequivalent center-symmetric minimal action
orbits that differ in the eigenphases of the spatial links.
However, this distinction is critical at finite volume only. The
above construction implies that the phases of the constant spatial
links of the final configuration can be chosen to all fall in the
interval $(-\pi/L, \pi/L]$, where $L$ is the spatial lattice
dimension in lattice units. In the limit $L\rightarrow \infty$,
the spatial links of the configuration all tend to unity. The
arbitrarily small deviations from unity can only be observed by
non-contractible Wilson loops that wrap around the whole {\it
spatial} extent of the lattice. These are not observables in the
infinite volume limit and a center-symmetric orbit of minimal
action in this sense is {\it unique}.

In the infinite volume limit at a finite temperature $T$ any
center-symmetric orbit with vanishing curvature can be represented
by a constant Abelian connection. Using\equ{symvac} and the
previous observation that spatial links of this representative
tend to unity for large spatial volume, this center-symmetric
Abelian background connection is,
\begin{eqnarray}\label{symV}
g \bar V_4^{(0)}=g a_4&=&T\,{\rm
diag}(\theta_1^{(0)},\dots,\theta_N^{(0)})
\nonumber\\
&=&2\pi T\,{\rm
diag}\left(\frac{-N+1}{2N},\frac{-N+3}{2N},\dots,\frac{N-3}{2N},\frac{N-1}{2N}\right)\nonumber\\
g \bar V_i^{(0)}=g a_i&=&0,\ {\rm for}\ i=1,2,3 .
\end{eqnarray}

Quadratic fluctuations about the center-symmetric configuration of
temporal links have been considered previously (see for instance
\cite{GrPiYa81,Neuberger83} and (for $N\rightarrow\infty$)
in\cite{Polchinski93}). We here will examine the perturbation
series about the configuration\equ{symV} to all orders in the
$1/N$ expansion of the free energy.

\subsection{Topological configurations that
break the center symmetry} If any Polyakov loop of a topological
configuration does {\it not} vanish, it necessarily belongs to a
multiplet of minimal action configurations. If $N$ is not prime,
the configuration may break a subgroup of $Z_N$ only. However, the
flat connection $V_4=0$ breaks the $Z_N$-group completely. It is
one of the $N$ Abelian  configurations of the form,
\begin{equation}\label{brokenV}
g\bar V_4^{(q)}=\frac{2\pi T q}{N}\,{\rm
diag}((1-N),1,\dots,1,1);\ \ q=1,2,\dots,N
\end{equation}
These configurations have degenerate eigenphases and one cannot
argue that the spatial links of such a minimal action
configuration are Abelian. Contrary to the center-symmetric case,
it is not clear that the index $q$ uniquely identifies a minimal
action orbit in the infinite volume limit.

The configurations of\equ{brokenV} have been studied
extensively\cite{Weiss81}. They correspond to minima of the free
energy at high temperatures $T$ when corrections proportional to
the coupling $g^2(T)$ are negligible. [However, the homogeneous
vacua of\equ{brokenV} do {\it not} solve the infrared problem of
the high-temperature expansion observed by Linde\cite{Linde80} --
the high-temperature phase probably\cite{GrPiYa81} can be
described by domains of such vacua with different index $q$.]

Minimal action configurations that break the center-symmetry to a
subgroup of $Z_N$ also can be constructed  for non-prime $N$. They
could play a r\^ole in the (perhaps rather complex) phase
structure of an $SU(N)$-model with non-prime $N$. Minimal action
solutions that break the $Z(N)$-symmetry correspond to
perturbative minima of the free energy.  They are not the minima
of the free energy in a center-symmetric (confining) phase.

\section{Large $N$ expansion in a center-symmetric background}
We are interested in the expansion of the free energy of a
$U(N)$-model at finite temperature for large values of $N$. We
shall argue that the model is in a confining phase as long as the
center-symmetric background is stable. [More specifically, the
free energy density $F$ of a $U(N)$ gauge theory expanded about
the center-symmetric background is ${\cal O}(N^0)$ rather than
${\cal O}(N^2)$ and ${\cal O}(N)$ as one expects when asymptotic
states form multiplets of the adjoint, respectively fundamental,
representation of $SU(N)$.]

The center-symmetric background of\equ{symV} is a \emph{maximum}
of the 1-loop free energy\cite{Weiss81}, whose minima are at the
configurations of\equ{brokenV} that spontaneously break the center
symmetry. It was recently found\cite{Diakonov04} that the
non-perturbative contribution from calorons\cite{Harrington78},
can make the minima of the 1-loop free energy unstable at low
temperatures. Near the deconfinement transition, calorons with
non-trivial holonomy have been observed by cooling $SU(2)$ and
$SU(3)$ lattice configurations\cite{Baal04}. However, classical
solutions of finite action generally are suppressed in the limit
of large $N$ ($g^2N$ finite). A semi-classical mechanism for
restoring the center symmetry thus appears unlikely at large $N$.

Lattice studies at relatively small $N$ see a distribution of
values for the Polyakov loop in the confining phase, rather than a
strong concentration near ${\cal L}(\bf x)=0$. This can also be
seen by studying the strong coupling expansion in a gauge where
all temporal links except those on the $x_4=0$ time slice are set
to unity. To leading order, the measure for the eigenphases
$\theta_j({\bf x})$ of\equ{diagonalU} in this case is given by the
Vandermonde determinant of the eigenvalues of the non-trivial
temporal link,
\begin{equation}\label{Vandermonde}
[dU]\rightarrow \prod_{i=1}^N d\theta_i
\prod_{i>j}\sin^2\left(\frac{\theta_i-\theta_j}{2}\right)\ .
\end{equation}
This measure is gauge invariant and vanishes when any two
eigenphases coincide\footnote{Degenerate configurations that are
invariant under a non-Abelian subgroup of $U(N)$ thus have
vanishing weight at strong coupling.}. It is maximal when the $N$
phases are evenly distributed over the circle $[0,2\pi]$. For
small values of $N$, the dependence on the eigenphases
of\equ{Vandermonde} is rather weak. However, for $N\sim\infty$ the
support of the measure\equ{Vandermonde} becomes restricted to the
immediate vicinity of the configuration\equ{symvac}. The strong
coupling limit of $U(N\sim\infty)$ lattice gauge theory thus is an
example for the more general conjecture\cite{Witten80} that
fluctuations are suppressed at large $N$.

Since its expectation vanishes, the usual factorization argument
fails for the Polyakov loop in a center-symmetric phase. However,
at strong coupling one can explicitly show\cite{Soshnikov99} that
the distribution of ${\cal L}=\tr U=\sum_j \exp(i\theta_j)$
converges to a standard normal in the limit $N\rightarrow\infty$.
The fact that the variance of the Polyakov loop does not grow with
$N$ implies that the standard deviation of the eigenphases is of
order $1/\sqrt{N}$ only. This also is apparent
from\equ{Vandermonde}. $<{\cal L}^2>=1$ furthermore is consistent
with usual $1/N$ counting, which suggests that the standdard
deviation of the eigenphases is ${\cal O}(1/\sqrt(N))$ in the
confining phase even when the strong coupling limit does not
apply.

The orbit described by\equ{symV} is invariant under spatial
translations and rotations and minimizes the free energy density
of lattice gauge theory in the strong coupling limit. It also is
the only center-symmetric candidate for a perturbative vacuum
orbit. Since fluctuations of the temporal links are expected to be
small at large $N$, we now consider the perturbative expansion
about the background of\equ{symV} at large $N$.

The Euclidean time derivative of a minimally coupled field in a
non-trivial representation of the group occurs through the
covariant derivative only. In the background of\equ{symV} the time
derivative of a field $\Phi$ of the adjoint representation thus is
replaced by,
\begin{equation}\label{tderadj}
\partial_4\Phi^a_b\rightarrow\bar{\cal D}_4\Phi^a_b=\partial_4\Phi^a_b + ig[a_4,
\Phi]^a_b=\partial_4\Phi^a_b + \frac{2\pi iT}{N}(a-b)\Phi^a_b;\
a,b=1,\dots,N\ .
\end{equation}

The time derivative of fields $\Psi$ in the fundamental
representation is similarly replaced by,
\begin{equation}\label{tderfund}
\partial_4\Psi^{a}\rightarrow\bar{\cal D}_4\Psi^a=\partial_4\Psi^{a} + i g
a_4^{ab}\Psi^b =\partial_4\Psi^{a} + \frac{2\pi
iT}{N}(a-\frac{N+1}{2})\Psi^{a};\ a=1,\dots,N\ .
\end{equation}

Physical correlation functions are colorless. All color indices
are summed over. At any finite temperature and for any
$N$\equ{tderadj} and\equ{tderfund} imply that we can associate a
discrete "color momentum",
\begin{equation}\label{colormom}
\xi(a)=\frac{2\pi T}{N}(a-(N+1)/2); \ \ a=1,\dots,N\ ,
\end{equation}
with a color index of the fundamental representation. For
sufficiently large $N$ one is tempted to replace sums over color
indices by integrals and neglect the error due to the fact that
$\xi(a)$ only takes discrete values,
\begin{equation}\label{integratecolor}
\sum_{a=1}^N\rightarrow \frac{N}{2\pi T}\int_{-\pi T}^{\pi T}
d\xi(a) + {\cal O}(1/N)\ .
\end{equation}
Note that color momentum is in the compact interval $[-\pi T,\pi
T]$ that does not depend on $N$. Loop integrals over color
momentum do {\it not} induce new UV-divergences. [Something rather
similar occurs in solid-state physics where momenta are restricted
to a single Brillioun-cell -- the associated space is an infinite
(periodic) lattice of points. There are no UV-divergences in this
case, since the smallest distance is the lattice spacing.] The
factors of $N$ (see \equ{integratecolor}) from the loop integrals
over color momentum can almost all be absorbed by redefining the
coupling,
\begin{equation}\label{lambda}
 g^2 N\rightarrow \lambda\ .
\end{equation}
Contrary to the $1/N$-expansion in the broken phase, the reduced
coupling $\lambda$ does depend on $N$. We argue in section~4.2
that the remaining dependence is logarithmic only.

In his seminal work on large $N$\cite{THooft74a} 't Hooft has
shown that the contribution of a connected vacuum diagram in
ordinary perturbation theory is proportional to a power of $N$
that depends on the difference in the number of color- and
momentum- loops of the diagram only. Using 't Hooft's double-line
notation one obtains a topological expansion of $U(N)$ gauge
theory in terms of the genus of perturbative diagrams. For a
background configuration of\equ{brokenV} that breaks the center
symmetry, the topological expansion coincides with an expansion in
powers of $1/N$ (in powers of $1/N^2$ when there are no
fundamental representations). In the broken phase, contributions
to the free energy of leading order in $N$ are given by planar 't
Hooft diagrams that have the topology of a two-sphere, $S_2$.

Some of the characteristics of the usual $1/N$ expansion are
retained by an expansion of the model about the center-symmetric
but $N$-dependent configuration of\equ{symV}. Since the background
is diagonal in color, one can still follow the color flow using 't
Hooft's double-line notation (see Fig.~1). \vspace{.5cm}

\epsfig{file=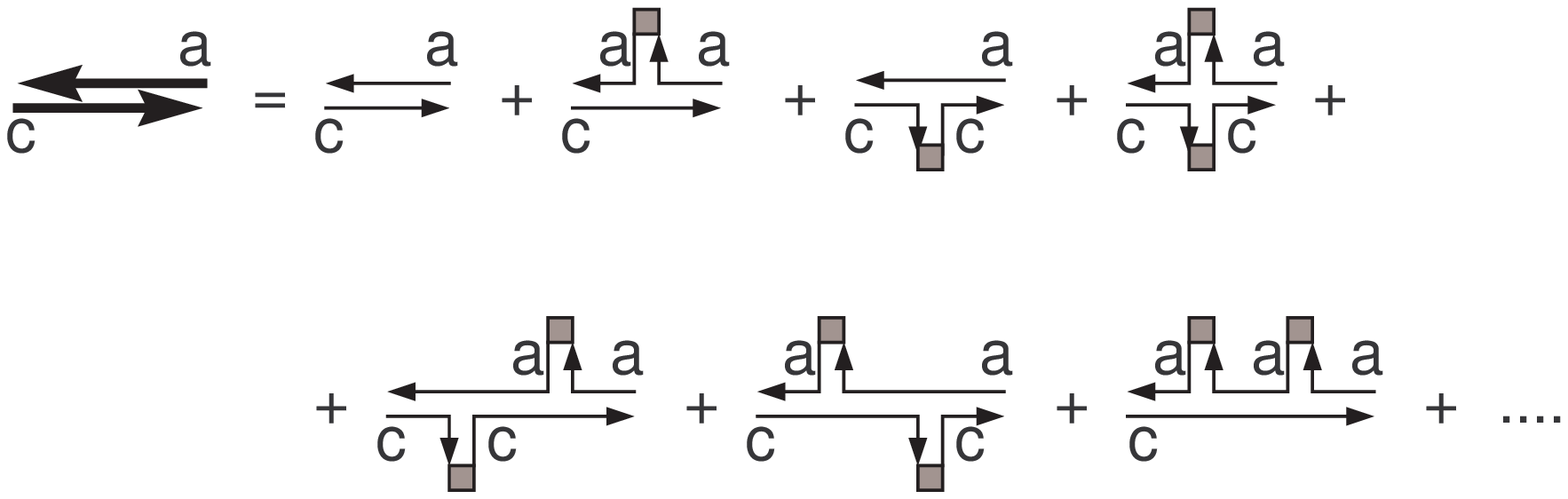,width=5truein}
 {\newline\small Fig.1: 'T
Hooft's double-line notation for the "dressed" gluon propagator:
insertions of the diagonal Abelian background of\equ{symV}
(depicted by shaded squares) do not change the color of a line ! }
\vspace{.5cm}

One therefore still has a topological expansion in the genus of
the 2-dimensional surfaces described by 't-Hooft diagrams.
However, {\it this topological expansion in general no longer
coincides with an expansion in $1/N$.} If one could neglect the
error due to the discreteness of color momentum
in\equ{integratecolor}, each color loop indeed would contribute a
factor of $N$ only. One then reaches the same conclusions about
the order of a diagram as in the broken case. However, due to the
discretization error, diagrams of a given genus in the topological
expansion may also contribute to higher orders of the $1/N$
expansion. The genus of a diagram thus only gives the {\it lowest}
(superficial) order in the $1/N$ expansion to which it may
contribute. This has some interesting consequences for the $1/N$
expansion of the free energy.

We show below that the contribution of planar diagrams to the free
energy of a $U(N)$ gauge theory without fundamental fields is of
order $1/N^2$ in the center-symmetric background of\equ{symV}.
[Although they did not specify the order in $1/N$, Gocksch and
Neri\cite{GoNe83} also found that planar diagrams do not
contribute at $N=\infty$.] The leading gluonic contribution to the
free energy is of order $N^0$ and given by 't~Hooft diagrams with
the topology of a torus.

\subsection{Planar $U(N)$ at finite temperature}
The flow of color momentum in planar diagrams is closely
associated with that of ordinary momentum. Consider gluonic
(vacuum) diagrams without external legs in the double-line
notation of 't~Hooft\cite{THooft74a}.  The number of momentum
loops, $L_p$, of a vacuum diagram with $E$ gluon propagators and
$V$ interaction vertices is,
\begin{equation}\label{momloops}
L_p=E-V+1\ .
\end{equation}
If the diagram is "planar", it has the topology of a
2-sphere\cite{THooft74a}, $S_2$. Gluon propagators are the edges
of cells and the Euler number, $\chi$, of a diagram is,
\begin{equation}\label{Eulernr}
\chi=V-E+L_c\ .
\end{equation}
Here $L_c$ is the number of faces, $V$ is the number of vertices
and $E$ is the number of edges of the complex. {'T}~Hooft's
double-line notation shows that the number of faces, $L_c$, is
just the number of independent traces over fundamental color
indices, that is the number of loops over color momentum.
\equ{momloops} and \equ{Eulernr} with $\chi(S_2)=2$ imply that,
\begin{equation}\label{loopcounting}
L_c=L_p+1\ {\rm for~planar~vacuum~diagrams} \ .
\end{equation}

As indicated in Fig.~2a, the loops over ordinary momentum can be
chosen to coincide with the color traces in planar diagrams. One
can enforce ordinary momentum conservation at each vertex by
writing the momentum of a gluon propagator as the {\it difference}
of two loop momenta associated with each face of the oriented
cells the propagator is an edge of. In vacuum diagrams one ends up
with just as many loop momenta as color traces. However, one of
these loop momenta amounts to an overall translation of all other
momenta and is redundant. We again arrive at\equ{loopcounting}.
\vspace{.5cm}

\epsfig{file=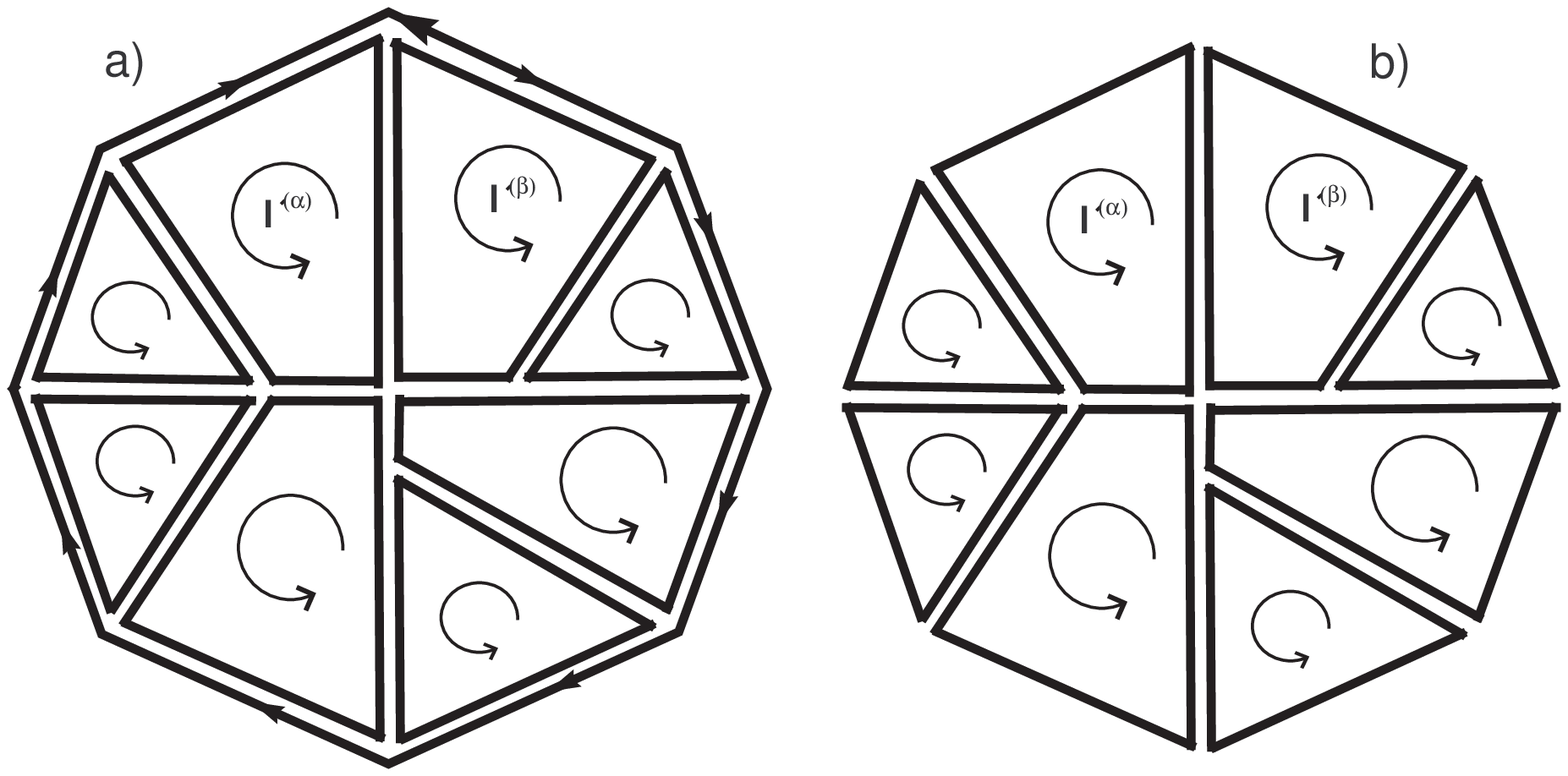,width=5truein}
\newline {\small\noindent Fig.2: a) A typical planar
gluonic vacuum diagram that superficially is of order $N^2$. The
flow of color {\it and} of ordinary momentum on each of the faces
$\alpha,\beta,\dots$ is given by composite loop momenta
$l^{(\alpha)}, l^{(\beta)}\dots$. The trace over color for the
perimeter loop results in a factor of $N$. b) A typical planar
vacuum diagram with one fundamental loop that superficially is of
order $N$. The flow of color {\it and} ordinary momentum again is
captured by composite loop momenta but there is no trace over
color only.} \vspace{.5cm}

 This association between color- and
momentum- loops in planar diagrams can be exploited. In
equilibrium at finite temperature, gluons are periodic fields in
Euclidean time with period $1/T$. Their Matsubara frequency
$\omega_n$ therefore is an integer multiple of the fundamental
frequency $2\pi T$,
\begin{equation}\label{mom4}
\o_n=2\pi T n,\ n\in {\bf Z}.
\end{equation}
We may enforce momentum conservation at a vertex (also at
fermion-gluon vertices) by writing the Matsubara frequency of a
gluon as the difference of the temporal components of two {\it
half-integer} loop momenta associated with the faces (say $\alpha$
and $\beta$) it is an edge of,
\begin{equation}\label{diffmom}
\o_{n-m}=k^{(\alpha)}_4(n)-k^{(\beta)}_4(m)= 2\pi T (
(n+1/2)-(m+1/2))=2\pi T (n-m)\ .
\end{equation}
For a planar vacuum diagram the loop momenta $k^{(\alpha)}_\m$ can
be chosen to run along the color loops. We thus can combine the
time-component of loop momentum $k^{(\alpha)}_4$ with the color
momentum $\xi^{(\alpha)}$ to the temporal component of a {\it
single} composite loop momentum $l_4^{(\alpha)}$,
\begin{equation}\label{fullmom}
l^{(\alpha)}_4(n,a)=k^{(\alpha)}_4(n)+\xi^{(\alpha)}(a)=\frac{2\pi
T}{N}( N n+a-1/2);\ n\in {\bf Z}, a=1,\dots,N\ .
\end{equation}
The conservation of the time component of ordinary loop momentum
{\it and} of color at a vertex, thus is equivalent to the
conservation of the integer $j=N n+a$, i.e. the time component of
composite loop momentum $l_4$. Note that the sum over the temporal
component of composite momentum extends over all half-integers and
that the temperature effectively is $T/N$ in planar $U(N)$. In
purely gluonic planar vacuum diagrams, every summation over a
composite loop index apart from one (the "peripheral" color loop)
is accompanied by a factor of $T$. For the peripheral loop of a
gluonic planar vacuum diagram the summation is over color only. It
amounts to a translation of all other composite loop momenta by a
half-integer between $1/2$ and $N+1/2$. This changes all other
summations over half-integer composite loop momenta to summations
over {\it integer} composite loop momenta and in addition yields
an overall factor of $N$ [since the expression for the diagram in
fact does not depend on finite shifts of all composite loop
momenta by integer multiples of $2\pi T/N$].

A planar gluonic vacuum diagram with $L_p$ momentum loops is of
perturbative order $(g^2)^{L_p-1}$ and is proportional to a factor
$N T^{L_p}$ due to $L_p$ summations over integer composite loop
momenta and the trace over color of the peripheral loop. For the
background of\equ{symV}, the regularized\footnote{The
superficially quartic ultraviolet divergence of the free energy
can be reduced to the superficially quadratic divergence of the
specific heat. It then is sufficient to regulate the spatial
integrals (see section~4.2). The severe infrared divergences of
perturbation theory observed by Linde\cite{Linde80} are absent in
the present case (see section~6).} planar contributions to the
free energy density, $ F_{S_2}(T) $, scale as,
\begin{equation}\label{scaling}
F_{S_2}(T,g^2,N)= N T f_{S_2}(T/N,\lambda)=T^4/N^2\tilde
f_{S_2}(\lambda)\ .
\end{equation}
After the UV-regularization is removed $\tilde f(\lambda)$ is a
dimensionless function of the reduced physical coupling
$\lambda(T/\Lambda)$, where $\Lambda$ is the appropriate
asymptotic scale parameter of the renormalization scheme (see
section~4.2).

\equ{scaling} shows that there is no contribution of order $N^2$
to the free energy from planar diagrams in the center-symmetric
background of\equ{symV}. This implies the absence of asymptotic
states in the adjoint representation of the group, that is of
(constituent) gluons, in center-symmetric $U(N)$. The free energy
of the model otherwise would have to be proportional to $N^2$, the
degeneracy of such a multiplet. The result also eliminates the
possibility of asymptotic states in higher dimensional
representations. \equ{scaling} suggests that the leading
contribution to the free energy of gluonic and center-symmetric
$U(N)$ is of order$ N^0$ and given by diagrams with the topology
of a doughnut $T_2$.

Although this is more or less what one would expect for the
confining phase of the model, some omissions and apparent
contradictions have to be addressed. Any explicit calculation
requires the specification of a gauge and an appropriate
regularization procedure. We have to show the existence of a gauge
that is compatible with the background of\equ{symV} and does not
invalidate the previous argument. We also still have to
verify\equ{scaling} for the (planar) contribution to the free
energy of $U(N)$ of order $\lambda^0$. This "1-loop" contribution
to the free energy is a Casimir energy that does not correspond to
an evaluation of vacuum diagrams like those discussed above. The
following sections support the above argument in important ways.

\section{Gauge Fixing and Renormalization}
\subsection{Background Gauge}
The background configuration of\equ{symV} is in the maximal
Abelian subgroup of $U(N)$ and a crucial point of the previous
argument was that all fields couple minimally to it. Covariant
Maximal Abelian gauges (MAG) satisfy this requirement and
furthermore can be defined\cite{Schaden99} on the
lattice\footnote{The lattice in this case is just a theoretical
framework for defining the regularized model, and not a very
convenient numerical tool.}. The Abelian Ward Identity of MAG
implies that the background $ga_\m$ does not renormalize in these
gauges\cite{Quandt98}. It therefore is sensible to set this
background connection proportional to the physical temperature $T$
in MAG.

However, the fact that MAG distinguishes between diagonal and
off-diagonal components of the connection gives rise to additional
vertices at which the color flow is {\it constrained}. In diagrams
containing such vertices, not all color loops are independent.
This leads to apparent modifications of the $1/N$-expansion and
complicates the $1/N$-counting considerably: due to cancellations,
gauge-invariant combinations of diagrams can be of different order
in $1/N$ than the connected diagrams are individually.

However, the free energy density of $U(N)$ is a gauge invariant
quantity and its expansion in $1/N$ should not depend on the
particular gauge. For the purpose of $1/N$-counting, background
gauges\cite{ZuKl75} in fact are much easier to use than covariant
MAG. Contrary to MAG one cannot define the BRST-symmetry of
background gauges on the lattice\cite{Neuberger87} since the
lattice gauge group is compact\cite{Schaden99}. But these are
renormalizable gauges that are well defined to all orders in
perturbation theory\cite{ZuKl75,Grassi96}. This suffices for our
purpose. Background gauges and MAG share the crucial properties
that the background $ga_\m$ does not renormalize\cite{ZuKl75,
Grassi96} and that it couples minimally to the fields. Since
background gauges are linear, they do not constrain the color flow
and do not change the $1/N$ counting of a diagram.

The background gauge in our case is defined by a gauge-fixing part
of the Lagrangian of the form,
\begin{equation}\label{LGF}
{\cal L}^{b.g.}_{GF}=\frac{1}{2\alpha} [{\bar{\cal D}}_\m
V^a_{\m\; b} ][{\bar{\cal D}}_\m V^b_{\m\; a} ]-{\bar
C}^b_a{\bar{\cal D}}_\m ({\cal D}_\m C)^a_b\ .
\end{equation}
${\bar{\cal D}}_\m$ and ${\cal D}_\m$ in\equ{LGF} are,
respectively, the background covariant derivative (with the
connection $ga_\m$ defined in\equ{symV}) and the ordinary
covariant derivative (with connection $g V_\m$). $C$ and ${\bar
C}$ denote the ghost and anti-ghost fields and $\alpha$ is the
gauge parameter. Upon shifting the gauge field $V_\m$ by the
constant and Abelian background $a_\m$, the premise that all time
derivatives occur as background covariant derivatives holds in
these gauges.

Apart from rigorously defining perturbative propagators and
introducing a set of adjoint ghost fields, there are no
constraints on the color summations in the background gauge fixing
of\equ{LGF}. These gauges therefore do not modify any of the
previous arguments with regard to the order in $N$ of a
perturbative diagram.

\subsection{Regularization and Renormalization}
Background gauges are renormalizable to all orders in perturbation
theory\cite{Grassi96}. We nevertheless have to show that the
previous scaling argument is not spoiled by the renormalization
procedure. Although the free energy density superficially diverges
quartically, the specific heat at constant volume ($C_{\cal
V}=-T\partial_T^2 F$) is only quadratically divergent. $C_{\cal
V}$ may, for instance, be regularized by analytic continuation in
spatial dimensions only.

The free energy density is recovered by integration of the
specific heat with the boundary conditions that the specific
entropy, $-\partial_T F$, and the free energy density, $F$, vanish
at $T=0$. This is equivalent to subtracting from the free energy
density any contribution that is linear in the temperature. For
$D=3-\eps$ spatial dimensions, a dimensionally regularized
perturbative contribution to the specific heat is of the form,
\begin{equation}\label{regularized}
C_{\cal V}(T,N,{\hat g}^2;\eps,\m)=T\partial_T^2 G(T,N,{\hat
g}^2;\eps,\m)\ ,
\end{equation}
where $G(T,N,{\hat g}^2;\eps)$ is the formal expression of the
vacuum graph in $D$ spatial dimensions, $\hat
g^2=g^2\m^{-\eps}=\hat \lambda/N$ is the renormalized
dimensionless coupling and $\m$ is the renormalization scale. The
diagrammatic argument of section~3.1 implies that the
contributions of planar gluonic vacuum graphs in the
center-symmetric background depend on $T$ and $N$ in the
particular combination,
\begin{equation}\label{scalingreg}
G_{S_2}(T,N,{\hat g}^2;\m,\eps)=N T f_{S_2}(T/N,{\hat
\lambda};\eps,\mu)=\frac{T^4}{N^2} {\tilde f}_{S_2}({\hat
\lambda};N\m/T ,\eps)\ .
\end{equation}
The subtraction of a constant term and of a term proportional to
$T$ from $G$ amounts to the subtraction from $f$ of a term
proportional to $N/T$ and of a $T$-independent constant. Possibly
divergent terms from planar vacuum diagrams that are proportional
to $N^2$ and $N$ thus do not contribute to the specific heat nor
to the free energy density.

Further, since the free energy is a physical quantity, $f({\hat
\lambda};N\m/T ,\eps)$ does not depend on the renormalization
point $\mu$. In the renormalization scheme (RS), $f({\hat
\lambda};N\m/T ,\eps\rightarrow 0^+)$ therefore is a function of
the renormalization group invariant effective coupling
$\lambda(T/\Lambda_{RS})$ only.

The free energies of center-symmetric planar $U(N)$ for different
$N$ are proportional only if the temperature is measured in terms
of an asymptotic scale parameter, $\Lambda_{RS}$, that does not
depend on $N$. To determine this finite renormalization, it is
sufficient to, for instance, demand that the deconfinement
temperature $T_d(N)$ of planar $U(N)$ be the same for all $N$. The
scaled free energy $N^2 F_{S_2}(T,N)$ then does not depend on $N$
at any temperature below $T_d$.

\equ{scalingreg} shows that the coupling $\lambda(N\m)$ is a
function of $N\m$ rather than of the renormalization point $\m$
only. Large values of $N$ correspond to large values of $\m$ --
and to weak coupling. For large $N$ this implies that,
\begin{equation}\label{ASP}
\lambda(\m N)\sim \frac{24\pi^2}{11\ln\frac{\m N}{\Lambda_{RS}}}\
.
\end{equation}
\equ{ASP} suggests that the confining phase can be explored
perturbatively at sufficiently large $N$.  This weak-coupling
confinement regime was first noticed by
Polchinski\cite{Polchinski93} while exploring the analogy between
string theory and large-N gauge theory. However, the background
of\equ{symV} is expected to be unstable for temperatures $T>T_d$.
Setting the renormalization mass $\m\sim T_d$ in \equ{ASP}, the
unstable regime corresponds to couplings $\lambda<\lambda_d$ with
\begin{equation}\label{critcoupling}
\lambda_d(N)\sim
\frac{24\pi^2}{11\ln\left(\frac{NT_d}{\Lambda_{RS}}\right)}\ .
\end{equation}
For any finite value of $N$, the phase transition occurs at a
(perhaps small) but nevertheless finite value of the coupling. An
asymptotic perturbative expansion thus is not possible in the
center-symmetric phase for any fixed value of $N$. Due to
\equ{critcoupling}, a perturbative evaluation of the (leading)
${\cal O}(N^0)$ contribution to the free energy could nevertheless
be reasonably accurate.

Since the usual $1/N$-expansion of $SU(N)$ is algebraic in $1/N$,
the logarithmic dependence of the coupling on $N$
in\equ{critcoupling} is somewhat unexpected. However, the
center-symmetric orbit of\equ{symvac} is described by a connection
that is itself $N$-dependent. This leads to a non-trivial
$N$-dependence of the momentum scale in planar diagrams. The usual
ultraviolet behavior of the model then results in a logarithmic
dependence on $N$ of the effective coupling. The perturbative
analysis of $U(N)$ gauge theory in the confining phase in this
sense becomes self-consistent at large $N$.

\section{Other contributions to the free energy of
center-symmetric $U(N)$ gauge theory} We saw that the contribution
from planar 't Hooft diagrams to the free energy in the
center-symmetric phase is of order $1/N^2$ only. 'T Hooft diagrams
with the topology of a torus may superficially contribute to the
free energy density in order $N^0$. To conclude that the free
energy of $U(N)$ indeed is of order $N^0$ in a center-symmetric
$1/N$ expansion we have to consider some remaining possibilities.

\subsection{No contributions to the free energy of ${\cal O}(N)$}
Fields in the fundamental representation of the group explicitly
break the center symmetry and superficially could give rise to
contributions to the free energy that are of order $N$. We will
see that there in fact are no such contributions in an expansion
about the center-symmetric background of\equ{symV}. The argument
is rather similar to the one employed in the gluonic case. Vacuum
diagrams that superficially are of order $N$ are planar diagrams
with one fundamental color loop only. [A sphere with a hole,
topologically a disc $D_2$.] A typical 't Hooft diagram of this
kind is shown in fig.~2b. We now have that $L_p=L_c$ and can
augment to composite loop momenta as before. The difference to
planar gluonic vacuum diagrams is the absence of an extra
perimeter loop over color only. This suppresses such contributions
by a factor of $N$ compared to the planar gluonic ones of fig.~2a.
The sums over the time-components of composite loop momenta now
extend over {\it half-integer} multiples of the fundamental
frequency $2\pi T/N$. The previous scaling argument shows that
such vacuum diagrams contribute to the free energy density in
order $1/N^3$:
\begin{equation}
F_{D_2}(T,g^2,N)= T f_{D_2}(T/N,\lambda)=\frac{T^4}{N^3}\tilde
f_{D_2}(\lambda(T/\Lambda))
\end{equation}
Below we explicitly find that this is also true for contributions
of order $\lambda^0$. There thus are no contributions of order
$N^2$ or $N$ in the expansion of the free energy density of a
$U(N)$ gauge theory about the center-symmetric background
of\equ{symV}. Since planar contributions to the free energy from
adjoint and fundamental fields vanish in the limit of large $N$,
the center-symmetric {\it planar} $U(N)$ model approaches a
topological theory without dynamical degrees of freedom.

\subsection{The free energy of center-symmetric $U(N)$ gauge
theory to order $\lambda^0$} The previous diagrammatic analysis
does not extend to the $1$-loop contribution to the free energy
density. We explicitly compute it for an $U(N)$ gauge theory with
$N_F$ Dirac fermions in the fundamental representation. The
relevant quadratic part of the Lagrangian is,
\begin{eqnarray}\label{freeL}
{\cal L}_0&=&\sum_{a,b=1}^N \Big[\frac{1}{4}(\bar{\cal D}_\m
V^{a}_{\n\;b} -\bar{\cal D}_\n V^{a}_{\m\;b})(\bar{\cal D}_\m
V^{b}_{\n\;a} -\bar{\cal D}_\n V^{b}_{\m\;a})
+\frac{1}{2\alpha}(\bar{\cal D}_\m
V^a_{\m\;b})(\bar{\cal D}_\n V^b_{\n\;a})\nonumber\\
&&+(\bar{\cal D}_\m \bar C^a_b)(\bar{\cal D}_\m C^b_a)\Big]
+\sum_{j=1}^{N_F} \sum_{a=1}^N \Big[\bar\Psi^j_{a}
\gamma_\m\bar{\cal D}_\m\Psi_j^a +i m_j
\bar\Psi^j_{a}\Psi_j^a\Big]\ .
\end{eqnarray}
In\equ{freeL} the $\gamma_\m$ are the hermitian Euclidean Dirac
matrices that satisfy $\gamma_\m \gamma_\n+\gamma_\n\gamma_\m={\bf
1}\delta_{\m\n}$. The time component of the background covariant
derivative $\bar{\cal D}_4$ for the fundamental and adjoint
representation is given in\equ{tderfund} and\equ{tderadj}
respectively ($\bar{\cal D}_\m =\partial_\m$ for spatial indices
$\m\neq 4$). The gluon- $(V_\mu)$ and ghost- $(C,{\bar C})$ fields
satisfy periodic boundary conditions in temporal direction whereas
the fermions $(\Psi_j,{\bar \Psi}_j)$ are anti-periodic.

Since the free energy does not depend on the gauge parameter, we
may for simplicity choose the Feynman-like gauge $\alpha=1$ to
compute it. For the constant background of \equ{symV}, the
eigenvalues of the operator $\bar {\cal D}_\m \bar{\cal D}_\m$ are
readily obtained and the functional integral over quadratic
fluctuations about this background can be formally performed. For
$D=3-\eps$ spatial dimensions, the regulated contribution to
$\partial_T^2F_0(T,N;\m,\eps)$, of the non-interacting model is:
\begin{eqnarray}\label{F01}
\partial_T^2 F_0(T,N;\m,\eps)=&\hspace{-10em}\partial_T^2
\frac{T}{2}\sum_{n=-\infty}^{\infty}\sum_{a=1}^N \int \frac{d^Dk
\m^\eps}{(2\pi)^D}\times \\
&\times\left\{\sum_{b=1}^N 2\ln\left[\frac{k^2+(2\pi T/N)^2(n N+
a-b)^2}{\m^2}\right]\right.\nonumber\\
&\hspace{3em}\left.- 4\sum_{j=1}^{N_F}\ln\left[\frac{k^2+(2\pi
T/N)^2(n N+ a-1/2)^2 +m_j^2}{\m^2}\right]\right\}\ .\nonumber
\end{eqnarray}
Noting that $n N+a$ ranges over all integers,\equ{F01} simplifies
to,
\begin{eqnarray}\label{F0}
\partial_T^2 F_0(T,N;\m,\eps)=&\partial_T^2
\frac{T}{2N}\sum_{n=-\infty}^{\infty} \int \frac{d^Dk
\m^\eps}{(2\pi)^D} \left\{2N^2\ln\left[\frac{k^2+(2\pi
T/N)^2n^2}{\m^2}\right]\right.\nonumber\\
&\left.- 4N\sum_{j=1}^{N_F}\ln\left[\frac{k^2+ (2\pi
T/N)^2(n-1/2)^2 +m_j^2}{\m^2}\right]\right\}\ .
\end{eqnarray}
This expression converges for $D<1$ spatial dimensions and thus is
at most quadratically divergent. Scale invariance of the
non-interacting model defined by\equ{freeL} implies the absence of
quadratic divergences in the massless case\cite{Sirlin03}.

The contribution to the free energy of a non-interacting massless
bosonic degree of freedom at temperature $T$ is finite and for
$D=3$ spatial dimensions is\cite{Kapusta89},
\begin{equation}\label{freeboson}
F_{boson}(T,m=0) = -\frac{T^4\pi^2}{90}\ .
\end{equation}
That from a non-interacting massive fermionic degree of freedom is
finite as well\cite{Kapusta89},
\begin{equation}\label{freefermion}
F_{fermion}(T,m)=\frac{m^2 T^2}{2\pi^2}
\sum_{n=1}^{\infty}\frac{(-1)^n}{n^2} K_2(n m/T)\ .
\end{equation}
In\equ{freefermion} $K_2(z)$ is the K-Bessel function normalized
so that for small arguments $K_2(|z|\sim 0)=2/z^2$. [Note that
$\frac{7}{8} F_{boson}(T,m=0)=F_{fermion}(T,m=0)\leq
F_{fermion}(T,m)\leq 0$.  The last inequality results because
$z^2K_2(z)$ is a monotonically decreasing function of its argument
on the positive real axis, with $z^2K_2(z)\leq 2$ for all $z\geq
0$. The contribution of massive fermions to the free energy
density is exponentially small for $T\ll m$.]

With the integration conditions that the free energy density and
the specific entropy vanish at zero temperature, (
$F_0(0,N)=\partial_T F_0(0,N)=0$), the specific heat completely
specifies the free energy density. One can read off the 1-loop
contribution to the free energy density from\equ{F0}: to lowest
order in the coupling, the free energy density of center-symmetric
$U(N)$ gauge theory at temperature $T$ is that of $2N^2$
non-interacting bosonic degrees of freedom and $4 N N_F$ (massive)
fermionic degrees of freedom {\it but at a temperature of $T/N$}.
Using\equ{freeboson} and \equ{freefermion} one has,
\begin{eqnarray}\label{F0sym}
F_0(T,N)&=2 N^2 F_{boson}(T/N,m=0)+4 N\sum_{j=1}^{N_F}
F_{fermion}(T/N,m_j)\nonumber\\
&= -\frac{T^4\pi^2}{45 N^2} +\frac{2T^4}{\pi^2
N^3}\sum_{j=1}^{N_F} \sum_{n=1}^{\infty}\frac{(-1)^n}{n^4}
\left(\frac{n N m_j}{T}\right)^2 K_2(n N m_j/T)\ .
\end{eqnarray}

We thus find the same behavior in $N$ for the 1-loop contribution
to the free energy density as for planar diagrams. We in fact used
the same arguments, combining color- and momentum- flow to a
single composite momentum $l_4(n,a)=(2\pi T/N)(n N+a)$. Although
the center-symmetric background effectively leads to an
$N$-dependent rescaling of the temperature, it is perhaps more
appealing to view this as an (for sufficiently large $N$) almost
complete cancellation of individual contributions to the free
energy (see appendix~A). Since background of\equ{symV} essentially
shifts frequencies by a fraction of the fundamental frequency
$2\pi T$ a partial cancellation can occur: for non-interacting
fields this phase shift effectively amounts to a change in
boundary conditions and the free energy density is sensitive to
this change\footnote{The contribution to the free energy density
of a bosonic degree of freedom satisfying anti-periodic boundary
conditions (corresponding to a shift of the frequency by $\pi T$)
for instance is positive.}. However, it is remarkable that this
cancellation is almost complete in planar $U(N)$ at large $N$.

\subsection{Contributions to the free energy of order $N^0$}
Center-symmetric {\it planar} $U(N)$ in the limit
$N\rightarrow\infty$ is devoid of physical degrees of freedom and
a topological model. Fortunately, the scaling arguments we used to
show the absence of contributions to the free energy of orders
$N^2$ and $N$ break down when there are {\it more} independent
momentum- than color- loops, that is when $L_p>L_c$. One then has
at least one loop momentum that cannot be augmented to a composite
momentum that includes the color flow. Such contributions to the
free energy density do not scale with $N$. The corresponding 't
Hooft diagrams either include more than one fermion loop or are
non-planar. The leading contribution of ${\cal O}(N^0)$ is given
by diagrams that topologically either are a torus or a sphere with
two holes [i.e. a disk with a hole]. These two classes of 't Hooft
diagrams correspond to contributions to the free energy of order
$N^0$ from non-interacting, colorless, asymptotic glueball- and
meson-states respectively. These are the stable asymptotic
states\cite{Witten79} at large $N$.

The fact that the free energy  is ${\cal O}(N^0)$ for sufficiently
large $N$ implies that in the center-symmetric background
of\equ{symV} {\it all} asymptotic states are color singlets. No
higher dimensional representation of the global color group
contributes. The center-symmetric expansion "confines" color in
this sense. [Note that this definition of confinement is somewhat
stronger than Wilson's screening criterion for static color
charges in the fundamental representation. The latter cannot be
applied in the presence of light meson states.]

To conclude that $U(N)$ gauge theory confines color charge at
sufficiently large $N$ one would have to show that the background
is {\it stable} against fluctuations in some (low) temperature
regime. The fact that the strong coupling expansion of lattice
gauge theory confines and is center-symmetric suggests that this
might be the case at sufficiently large effective coupling
$\lambda(T/\Lambda)>\lambda_d$. To further conclude that the more
realistic $SU(3)$-model confines color charge at low temperatures
one in addition has to show that there is no (deconfining) phase
transition at some finite $N>3$. Neither of these issues will be
discussed further here. Let us instead look at some interesting
aspects of the previous analysis.

\section{Infrared-Finite Perturbation Theory at Finite Temperature}
The suppression of planar contributions to the free energy density
shows that color momentum is essential. For a background
like\equ{brokenV} that breaks the center symmetry (and corresponds
to vanishing color momentum $\xi$), Linde observed\cite{Linde80}
that the perturbation series of a non-Abelian gauge theory is
infrared divergent at any finite temperature. The most infrared
divergent vacuum diagrams are all {\it planar} and superficially
are of order $N^2$. The center-symmetric background of\equ{symV}
provides an effective infrared cutoff of order $2\pi T/N$ for all
the coset excitations. Since it appears via the covariant
derivative, this infrared cutoff is {\it not} entirely equivalent
to an effective gluon mass. Unlike an effective mass, it does not
regulate the Abelian sector of the model in the infrared. The
gauge bosons of an Abelian $U(1)^N$-model on the other hand do not
interact directly, and the infrared behavior of such models is
regular when all charged fields are massive\cite{Kinoshita62}.
Even though the option of massive off-diagonal fields is not
available for a $U(N)$ gauge theory, the center-symmetric
background of\equ{symV} provides an infrared cutoff that works
rather similarly: it shifts the infrared singularity of any coset
field propagator from $\bf k^2=0$ to ${\bf k}^2=-4\pi^2 T^2
j^2/N^2$ for some integer $N/2>j>0$. Note that although some of
these "masses" are rather small at large $N$, they {\it do not}
depend on the coupling $\lambda$.

In the center-symmetric background of\equ{symV} the perturbation
series thus is free of infrared divergences without resummation.
Of course, when the effective coupling is sufficiently small (at
high temperatures) this background presumably is not
stable\cite{Weiss81} (see below). The center-symmetric background
of\equ{symV} cures the pervasive infrared problem of the
perturbative expansion at low temperatures only. Although the
perturbation series may not converge in this regime, its mere
existence to all orders does define the model formally. The
regularization of perturbative infrared divergences by the
center-symmetric background, however,  reshuffles contributions to
the $1/N$ expansion of the free energy. It does so in a manner
that is consistent with confinement in this phase.

\section{Stability and (Veneziano's) Vector Ghosts}
The result that planar $U(N)$ gauge theory practically has no
degrees of freedom at large $N$, implies that center-symmetric
{\it planar} $SU(N)$, although devoid of colored asymptotic
states, is {\it not} a thermodynamically stable model.
Center-symmetric $SU(N)$ gauge theory nevertheless can be a
perfectly good physical model because the subset of planar
diagrams does {\it not} give the {\it leading} contribution in
$1/N$. There then is no reason why this subset of diagrams should
define a thermodynamically viable physical model. Planar diagrams
are generated by the Cuntz algebra\cite{GoGr95} rather than by a
bosonic or fermionic one. There is no proof that such a field
theory is thermodynamically stable.

The instability of center-symmetric {\it planar} $SU(N)$ follows
immediately from the previous result for the planar $U(N)$ model
without fundamental fields. The color singlet "photon" decouples
in this case and the free energy density of $U(N)$ is just that of
the corresponding $SU(N)$ model and of a free photon.
\equ{freeboson} together with the previous result for $U(N)$
implies that the free energy density of center-symmetric planar
$SU(N)$ is,
\begin{equation}\label{SUNfree}
F_{S_2}^{SU(N)}(T)=\frac{T^4\pi^2}{45} + {\cal O}(1/N^2)\ .
\end{equation}
The {\it positive} contribution to the free energy of
center-symmetric $SU(N)$ is of ${\cal O}(N^0)$ and can be
interpreted as due to a massless, color-singlet {\it vector ghost}
that compensates the degrees of freedom of the massless,
color-singlet "photon" of center-symmetric planar $U(N)$.

Veneziano\cite{Veneziano79} has shown that a massless
color-singlet vector ghost in planar gluonic $SU(N)$ could
saturate the axial Ward Identities and solve the $U_A(1)$-problem
at large $N$. \equ{SUNfree} is evidence for the existence of a
vector-ghost in the confining phase of the planar model. Whether
this vector-ghost couples to the axial current in the manner
Veneziano suggests, cannot be determined from the free energy. To
have a viable confining phase, the vector-ghost of the planar
model would have to either decouple by itself (as all states in
the planar truncation do) or be part of a
BRST-quartet\cite{Kawarabayashi80} that does not contribute to the
free energy.

As discussed in the previous sections, there are additional
contributions to the free energy density of ${\cal O}(N^0)$ in the
center-symmetric phase that are described by non-planar diagrams.
These non-planar contributions to the free energy density depend
on the effective coupling $\lambda$. It is at least conceivable
that massless bound states form when $\lambda>\lambda_d$ that
complete the BRST quartet and compensate the contribution to the
free energy of the vector ghost. Since the vanishing of ghost
contributions to the free energy is necessary for the stability of
a center-symmetric phase, the critical coupling at which this
occurs is a lower bound for $\lambda_d$.

The fact that the non-interacting (Casimir) part of the free
energy density of $SU(N)$ is positive (from\equ{F0sym}
with\equ{freeboson}) for all $N\geq 2$, implies that the
center-symmetric phase is not stable at small effective coupling
$\lambda(T/\Lambda)\sim 0$. This is consistent with the
expectation that the center symmetry is broken for $T>T_d>0$.

When non-planar contributions to the free energy of ${\cal
O}(N^0)$ are included, $SU(N)$ could be thermodynamically stable
in the center-symmetric phase at sufficiently large effective
coupling $\lambda(T/\Lambda)>\lambda_d$.

\section{Discussion and Conclusion}
The configuration of\equ{symV} is an absolute minimum of the
classical action of unbroken $U(N)$ gauge theory. It is invariant
under the discrete global $Z_N$ center-symmetry of the Yang-Mills
action. The description of a center-symmetric orbit of vanishing
curvature is {\it unique} in the infinite volume limit of a
periodic lattice. It is unique even at finite volume if the
spatial topology is that of a three-sphere, since the only
non-contractible Wilson loops in this case are the Polyakov loops
in temporal direction.

Although the background of\equ{symV} is an absolute minimum of the
classical Yang-Mills action, it is not an absolute minimum of the
free energy density at all temperatures. Due to its symmetries,
this orbit is always an extremum of the free energy, but to lowest
order of perturbation theory  this extremum is a maximum. To the
extent that higher order perturbative corrections are negligible,
the center symmetry is broken for sufficiently small effective
coupling, that is at sufficiently high temperatures\cite{Weiss81}.
The strong-coupling expansion of lattice gauge theory suggests
that a center-symmetric phase is thermodynamically preferred at
low temperatures when the effective coupling is sufficiently
strong.

An expansion in $1/N$ could be an appealing alternative to the
strong coupling expansion in this non-perturbative regime. The
center-symmetry of the gluonic sector is preserved in every order
of the perturbative expansion about the background configuration
of\equ{symV}. Certain qualitative conclusions about the
$1/N$-expansion of the free energy can be obtained by examining
this perturbative series. There are no contributions to the free
energy of order $N^2$ or $N$ at any perturbative order. For large
$N$, a center-symmetric planar truncation of $U(N)$ approaches a
topological field theory without dynamical degrees of freedom.
This confirms the result of Gocksch and Neri\cite{GoNe83} that the
free energy of lattice gauge theory in the planar limit does not
depend on the temperature (and therefore vanishes) at large $N$.
The leading contribution is of order $N^0$ and due to vacuum
diagrams that could represent the free energy of color singlet
quark-antiquark mesons and glueballs.

This is as one expects for the confining phase of a $U(N)$ gauge
theory. Perhaps more significant is that perturbative calculations
at finite temperature in principle are feasible in the background
of\equ{symV}. The severe infrared divergences of ordinary
perturbation theory observed by\cite{Linde80} do not occur in this
center-symmetric expansion. The reduced coupling $\lambda $
furthermore becomes weak in the confining phase for sufficiently
large $N$ (see section~4.2). Some aspects of confinement therefore
may be accessible in a perturbative framework\cite{Polchinski93}.
The perturbative analysis of the model in the vicinity of the
center-symmetric minimal action orbit in this sense is
(self)consistent. However, the existence of a phase transition at
a perhaps very small but nevertheless finite value of the coupling
restricts the accuracy of a perturbative analysis in the confining
phase at any fixed value of $N$. More quantitative results perhaps
can be expected from summing classes of perturbative
contributions.

Planar $SU(N)$ turns out to be thermodynamically unstable at
sufficiently large $N$ due to a massless color-singlet
vector-ghost. This is not an artefact of our treatment but rather
a consequence of the fact that the $U(1)$-photon decouples and the
free energy of planar $U(N)$ in the confining phase vanishes as
$N\rightarrow \infty$. We speculate that the vector ghost couples
to the axial current in the manner conjectured by
Veneziano\cite{Veneziano79}. It then should be part of a
BRST-quartet\cite{Kawarabayashi80} that, as a whole, does not
contribute to the free energy. This could be the case at
temperatures $T<T_d$ when other contributions to the free energy
density of ${\cal O}(N^0)$ are included.

Although the center symmetry of the action is explicitly broken by
fields in the fundamental representation, they do not contribute
in order $N$ to the free energy in an expansion about the
center-symmetric background of\equ{symV}. Spontaneous chiral
symmetry breaking could perhaps be investigated in this
background: to lowest order in the coupling and for large $N$
several fermionic degrees of freedom are almost zero modes in the
background of\equ{symV}. Whether this is sufficient to trigger
spontaneous chiral symmetry breaking has not been explored. If so,
the thermodynamic instability of the background for $T>T_d$ would
imply that the chiral- and deconfinement-transition temperatures
coincide.

\vspace{0.5cm}

{\bf \noindent Acknowledgements}
\newline\noindent I am very much indebted to L.~Spruch, D.~Zwanziger
and L.~Baulieu for moral and intellectual support and thank my
colleagues at Rutgers University for making this investigation
possible. I thank D.~Kabat for helpful suggestions in the early
stages of this work and H.~Neuberger for drawing my attention to
Eguchi-Kawai models.

\appendix
\section{Casimir contributions to the free energy
of center-symmetric $U(N)$}

We here calculate contributions to the free energy density in the
center-symmetric background of\equ{symV} to zeroth order in
perturbation theory without recourse to the scaling argument. The
calculation explicitly shows that contributions of individual
degrees of freedom cancel.

Consider the regularized expression\footnote{For simplicity and to
easily include fermions, the following computation uses
dimensional regularization. The gluonic contribution to the free
energy of $SU(N)$ has also been computed in lattice
regularization\cite{Neuberger83}. Evaluating Neuberger's lattice
result confirms that this gluonic contribution to the free energy
is $O(1/N^2)$ at the center-symmetric background and sub-leading
for sufficiently large $N$.} for the free energy density given
by\equ{F01}. The free energy density to lowest order in the
coupling can be decomposed,
\begin{eqnarray}\label{F03}
F_0(T,N)=2\sum_{a,b=1}^N I(T,(a-b)/N;0)-
4\sum_{j=1}^{N_F}\sum_{a=1}^N I(T,(a-1/2)/N;m_j)\ ,
\end{eqnarray}
into individual contributions $I(T,\delta;m)$ that depend on the
phase $\delta$ and mass $m$ associated with a particular degree of
freedom.  $I(T,\delta;m)$ is formally given by,
\begin{equation}\label{defI}
I(T,\delta;m)=\lim_{D\rightarrow 3^-} \frac{T}{2}
\sum_{n=-\infty}^{\infty}
\int\frac{d^Dk\m^{3-D}}{(2\pi)^D}\ln(k^2+m^2+4\pi^2
T^2(n+\delta)^2)
\end{equation}
The expression of\equ{defI} is not well defined. However, making
use of the fact that the free energy density $F$ {\it and} the
specific entropy $-\partial_T F$ vanish at $T=0$, it suffices to
obtain the second derivative $\partial_T^2 I$. Explicitly taking
the derivatives in\equ{F01}, the expression for $\partial_T^2
I(T,\delta;m)$ may be written,
\begin{eqnarray}\label{I}
\partial_T^2 I(T,\delta;m)&=&\lim_{D\rightarrow 3^-}
\sum_{n=-\infty}^{\infty}
\int\frac{d^Dk\m^{3-D}}{(2\pi)^D}\frac{4\pi^2 T
(n+\delta)^2}{k^2+m^2+4\pi^2 T^2(n+\delta)^2}\\ &&\times
\left[1+\frac{2(k^2+m^2)}{k^2+m^2+4\pi^2
T^2(n+\delta)^2}\right]\nonumber\\ &=&-\lim_{D\rightarrow 3^-}
\partial_T^2 T \int_0^\infty
\frac{d\lambda}{2\lambda}\sum_{n=-\infty}^{\infty}\int\frac{d^Dk\m^{3-D}}{(2\pi)^D}
e^{-\lambda(k^2+m^2+4\pi^2 T^2(n+\delta)^2)}\nonumber
\end{eqnarray}
[Note that the final (finite) result does not depend on the
renormalization point $\m$. The latter was introduced to have a
free energy density with the canonical dimension.] We next
evaluate the momentum integrals in the last expression of\equ{I},
\begin{eqnarray}\label{II}
\partial_T^2 I(T,\delta;m)&=&-\lim_{D\rightarrow 3^-}
\partial_T^2 T \int_0^\infty
\frac{d\lambda}{2\lambda}\frac{\m^3e^{-\lambda
m^2}}{(4\pi\lambda\m^2)^{D/2}}\sum_n e^{-4\pi^2\lambda
T^2(n+\delta)^2}.
\end{eqnarray}

To separate the summation over the integers $n$ from the
dependency on $\delta$ it is convenient to use the
Fourier-representation of the Gaussian :
\begin{eqnarray}\label{Gaussian}
\sum_n e^{-4\pi^2\lambda T^2(n+\delta)^2}&=&\int_{-\infty}^\infty
\frac{dp}{T \sqrt{4\pi\lambda}} e^{-p^2/(4 T^2\lambda)}
e^{2\pi i p \delta } \sum_n e^{2\pi i p n}\nonumber\\
&=& \sum_n \frac{e^{2\pi i
n\delta}}{T\sqrt{4\pi\lambda}}e^{-n^2/(4 T^2\lambda)}\\
&=&\frac{1}{T\sqrt{4\pi\lambda}}\left\{1+2\sum_{n=1}^\infty
\cos(2\pi n\delta) e^{-n^2/(4 T^2\lambda)}\right\}\ .\nonumber
\end{eqnarray}
[Note that the second expression for the sum is dual to the
original one in the sense that the "radius" of the temporal
direction has been inverted $4\lambda T^2\rightarrow 1/(4\pi^2
\lambda T^2)$.] Substituting the last expression for the sum
into\equ{II} and noting that the constant term in the braces
of\equ{Gaussian} does not survive differentiation, one finds that
in $D=3$ dimensions,
\begin{eqnarray}\label{Ifin}
I(T,\delta;m)&=& -\sum_{n=1}^\infty \cos(2\pi
n\delta)\int_0^\infty \frac{d\lambda}{\lambda}\frac{e^{-\lambda
m^2-n^2/(4
T^2\lambda)}}{(4\pi\lambda)^2}\nonumber\\
&=& -\sum_{n=1}^\infty \frac{\cos(2\pi n\delta)}{2\pi^2
}\left(\frac{T m}{n}\right)^2 K_2(n m/T)\ .
\end{eqnarray}
The integration constants in\equ{Ifin} have been determined so
that the free energy density and the specific entropy vanish at
$T=0$.

The result of\equ{Ifin} can be checked in various limits: the free
energy density of a non-interacting bosonic degree of freedom
satisfying periodic boundary conditions is obtained with
$\delta=0$; $\delta=1/2$ corresponds to the (positive) free energy
of a non-interacting bosonic degree of freedom satisfying
anti-periodic boundary conditions, etc.  One evidently can achieve
some cancellation in the total free energy by mixing several
degrees of freedom satisfying different boundary conditions. In
the center-symmetric background of\equ{symV}, the gluons
effectively satisfy different boundary conditions. If the
fundamental and anti-fundamental color indices of the gluon (in
the background gauge) are $a$ and $b$, the corresponding shift of
the Matsubara frequency is $2\pi T(a-b)/N$. Note that diagonal
Abelian degrees of freedom (with $a=b$) do not suffer a phase
shift and that all bosonic degrees of freedom are $N$-periodic.

The gluonic part of the free energy density in\equ{F03} is found
by evaluating,
\begin{eqnarray}\label{cancelglue}
2\sum_{a,b=1}^N \cos(2\pi n (a-b)/N) &=& \sum_{a,b=1}^N e^{2\pi i
n(a-b)/N} + c.c.\nonumber\\
&=& 2 \left|\sum_{a=1}^N e^{2\pi i a n/N}\right|^2\nonumber\\
&=& 2 N^2 \sum_k \delta_{n,kN}\ .
\end{eqnarray}
Here $\delta_{n,kN}$ is the Kronecker symbol (one when $n$ is an
integer multiple of $N$ and zero otherwise). Semiclassically,
contributions to the free energy density from gluonic paths with
$(n\ {\rm mod}\ N)\neq 0$ windings thus cancel completely.
Using\equ{Ifin} (for $m\rightarrow 0$) and\equ{cancelglue} the
gluonic contribution in\equ{F03} is,
\begin{eqnarray}
2\sum_{a,b=1}^N I(T,(a-b)/N;0)&=&-2T^4\sum_{a,b=1}^N
\sum_{n=1}^\infty\frac{\cos(2\pi n(a-b)/N)}{\pi^2 n^4}\nonumber\\
&=&-T^4\sum_{k=1}^\infty\frac{2 N^2}{\pi^2 (kN)^4}=
-\frac{T^4\pi^2}{45 N^2}\ .
\end{eqnarray}
This verifies the scaling argument for the first term
of\equ{F0sym}. The fermionic contribution in\equ{F0sym} is
similarily obtained from\equ{Ifin} and\equ{F03}. We have that,
\begin{eqnarray}\label{cancelquark}
\sum_{a=1}^N \cos(2\pi n (a-1/2)/N) &=& \frac{1}{2}\sum_{a=1}^N
e^{2\pi i
n(a-1/2)/N} + c.c.\nonumber\\
&=& N \sum_k (-1)^k \delta_{n,kN} \ .
\end{eqnarray}
Using\equ{cancelquark} and\equ{Ifin} in\equ{F03} gives the second
term of\equ{F0sym}.

\end{document}